\documentclass[12pt]{article}

\usepackage{epsf}
\usepackage[dvips]{graphicx}

\begin{document}

\begin{center}
{\bfseries STRUCTURE FUNCTION $F_L$ AT FIXED  $W$ IN THE 
$K_T$-FACTORIZATION APPROACH}

\vskip 5mm

A.V. Kotikov$^{1}$, A.V. Lipatov$^{2}$ and N.P. Zotov$^{2 \dag}$

\vskip 5mm

{\small
(1) {\it
BLThPh,  Joint Institute for Nuclear Research,
141980 Dubna, Russia}
\\
(2) {\it
Skobeltsyn Institute of Nuclear Physics,
Lomonosov Moscow State University\\
119992 Moscow, Russia}
\\
$\dag$ {\it
E-mail: zotov@theory.sinp.msu
}}
\end{center}

\vskip 5mm

\begin{center}
\begin{minipage}{150mm}
\centerline{\bf Abstract}
The results for structure function $F_L$, obtained in the
$k_T$-factorization and collinear approaches, are
compared with recent H1 experimental data at fixed $W$ values.
\end{minipage}
\end{center}

\vskip 10mm

\section{Introduction}

The longitudinal structure function (SF) $F_L(x,Q^2)$
is a very sensitive QCD characteristic and is directly connected to the
gluon content of the proton.
It is equal to zero in the parton model with spin$-1/2$ partons
and has got nonzero values in the framework of perturbative Quantum
Chromodynamics.
The perturbative QCD, however, leads to a quite controversal results.
At the leading order (LO) approximation $F_L$ amounts to about $10\div 20
\%$ of the corresponding $F_2$ values at large
$Q^2$ range and, thus, it has got quite large contributions at low $x$ 
range.
The next-to-leading order (NLO) corrections to the longitudinal
coefficient function are large and negative at small $x$
\cite{Neerven}-\cite{Rsmallx}
and can lead to negative $F_L$ values at
low $x$ and low $Q^2$ values (see \cite{Rsmallx,Thorne}).
Negative $F_L$ values demonstrate a limitations of the applicability of
perturbation theory and the necessity of a resummation procedure,
that leads to coupling constant scale higher than $Q^2$
(see \cite{Rsmallx}, \cite{DoShi}-\cite{Salam}).

The experimental extraction of $F_L$ data requires a
rather cumbersome procedure, especially at small values of $x$.
Recently, however,  there have been presented
new precise preliminary H1 data  \cite{H1rev} on the longitudinal SF
$F_L$,
which have probed the small-$x$ region $ 10^{-5} \leq x \leq 10^{-2}$.

In Ref.~\cite{KLZ}
 the standard perturbative QCD formulas and also the so called
$k_T$-factorization approach \cite{CaCiHa}
based on Balitsky-Fadin-Kuraev-Lipatov (BFKL) dynamics \cite{BFKL}
are used for the analysis of the above data.  Here we present
the main results of our analysis.

In the framework of the $k_T$-factorization approach  a study of the
longitudinal SF $F_L$ has been done firstly in Ref. \cite{CaHa}.
We follow a more phenomenological approach \cite{KLZ1}(see also~\cite
{Blumlein93,BaKwSt}),
where we analyzed $F_L$ data in a broader range at small $x$, using
the different parameterizations
of the unintegrated gluon distribution function $\Phi_g(x,k^2_{\bot})$
(see Ref. \cite{Andersson}).

\section{Theoretical framework}

The  unintegrated gluon distribution $\Phi_g(x,k^2_{\bot})$
($f_g$ is the (integrated) gluon distribution in the proton multiplied
by $x$ and $k_{\bot}$ is the transverse part of the gluon 4-momentum
$k^{\mu}$)
 \begin{eqnarray}
f_{g}(x,Q^2) ~=~ \int^{Q^2}dk^2_{\bot}
\, \Phi_g(x,k^2_{\bot})
~~~~~\mbox{(hereafter }
~k^2=-k^2_{\bot} \mbox{)}
\label{1}
 \end{eqnarray}
is the basic dynamical quantity in
the $k_T$-factorization approach
It satisfies the BFKL equation \cite{BFKL}.

Then, in the $k_T$-factorization
the SF $F_{2,L}(x,Q^2)$ are driven at small $x$ primarily
by gluons and are related in the following way to
$\Phi_g(x,k^2_{\bot})$:
\begin{eqnarray}
F_{2,L}(x,Q^2) ~=~\int^1_{x} \frac{dz}{z} \int^{Q^2}
dk^2_{\bot} \sum_{i=u,d,s,c} e^2_i
\cdot \hat C^g_{2,L}(x/z,Q^2,m_i^2,k^2_{\bot})~ \Phi_g(z, k^2_{\bot}),
 \label{d1}
\end{eqnarray}
where $e^2_i$ are charge squares of active quarks.

The functions $\hat C^g_{2,L}(x,Q^2,m_i^2,k^2_{\bot})$
can be regarded as  SF of the
off-shell gluons with virtuality $k^2_{\bot}$ (hereafter we call them
{\it hard structure functions } by analogy with similar
relations between cross-sections and hard
cross-sections). They are described by the sum of the quark
box (and crossed box) diagram contribution to the
photon-gluon interaction
(see, for example, Fig. 1 in \cite{KLZ1}).

Notice that the $k^2_{\bot}$-integral in Eqs. (\ref{1}) and (\ref{d1})
can be divergent at lower limit,
at least for some parameterizations of $\Phi_g(x,k^2_{\bot})$.
To overcome the problem we change the low $Q^2$ asymptotics of
the QCD coupling constant within hard structure functions.
We applied the ``freezing'' procedure \cite{NikoZa},which contains the 
shift $Q^2 \to Q^2 + M^2$, where $M$ is an additional
scale, which strongly modifies the infrared $\alpha_s$ properties.
For massless produced quarks, $\rho$-meson mass $m_{\rho}$ is usually
taken as the $M$ value, i.e. $M=m_{\rho}$.
In the case of massive quarks with  mass $m_i$, the $M=2m_{i}$ value
is usually used.For the unintegrated gluon distribution
$\Phi (x, k^2_{\bot}, Q_0^2)$ we use the so-called
Blumlein's parametrization (JB) \cite{Blumlein}.
Note that there are also several other popular parameterizations, which
give quite similar results  excepting,
perhaps, the contributions from the small $k_{\bot}^2$-range:
$k_{\bot}^2 \leq 1$ GeV$^2$ (see Ref. \cite{Andersson}).

The JB parametrization depends strongly on the Pomeron intercept value.
In different models the Pomeron intercept
has different values. So, in
our calculations we apply the H1 parameterization \cite{H1slope},
which are in good agreement with perturbative QCD.

We calculate the SF $F_L$ as the sum of two types of contributions -
the charm quark one $F^c_L$ and the light quark one $F^{l}_L$:
\begin{eqnarray}
F_L  ~=~ F^{l}_L + F^{c}_L.
\end{eqnarray}
For the $F^{l}_L$ part we use the massless limit of hard SF (see
\cite{KLZ1}).
We always use $f=4$ in  our fits, because our results depend very
weakly on the exact $f$ value.

\section{Numerical results}

\begin{figure}[!t]
\vspace*{7.0cm}
\begin{center}
\includegraphics{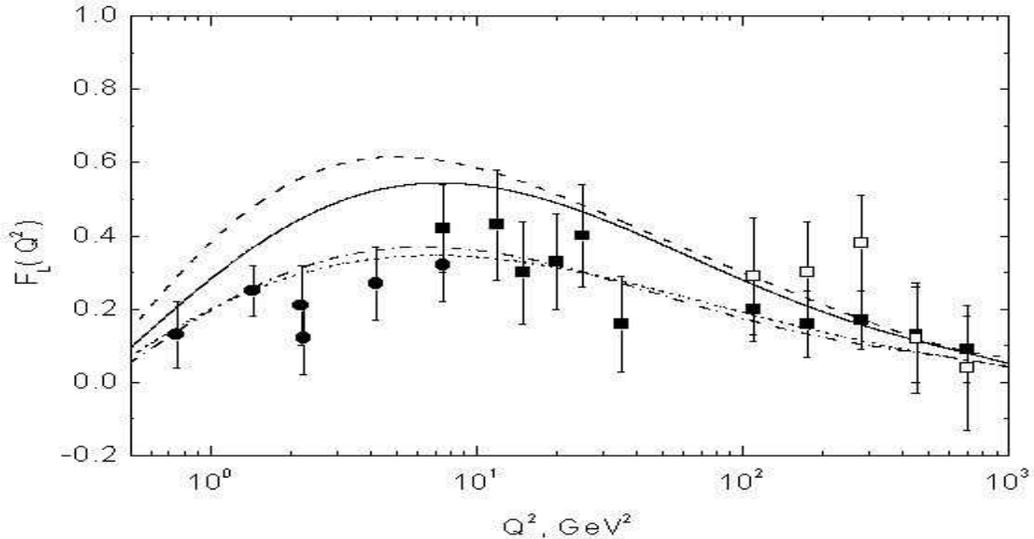}
\vspace*{.3cm}
\caption[*]{$Q^2$ dependence of $F_L(x, Q^2)$ (at fixed $W$ = 276 GeV).
The experimental points are from \cite{H1rev}.
Solid curve is the result of the $k_T-$factorization approach,
dashed, dash-dotted and dotted curves - the results of the collinear
LO, NLO and LO  with $\mu^2 = 127Q^2$ calculations, respectively.
}
\vspace*{-.3cm}
\end{center}
\end{figure}

\begin{figure}[!t]
\vspace*{7.0cm}
\begin{center}
\includegraphics{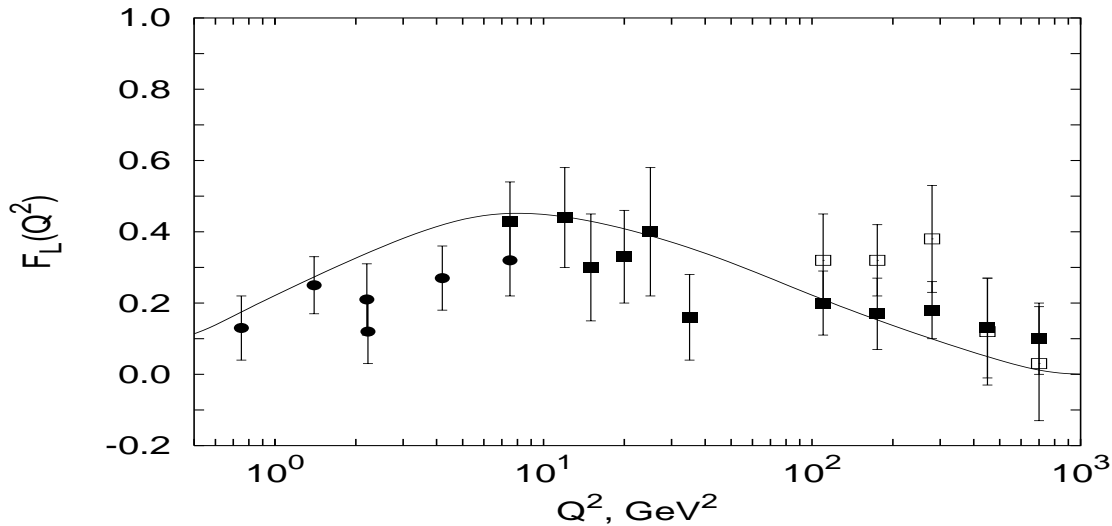}
\vspace*{-.2cm}
\caption[*]{$Q^2$ dependence of $F_L(x, Q^2)$ (at fixed $W$ = 276 GeV).
The experimental points are as in Fig. 1.
Solid curve is the result of the $k_T-$factorization approach
 with the GLLM unintegrated gluon distribution from~\cite{LL}.}
\end{center}
\end{figure}

In Fig. 1 we show  the SF $F_L$ with ``frozen''
coupling constant as a function of $Q^2$ for fixed
$W$  in comparison with H1 experimental data \cite{H1rev}.
The $k_T$-factorization results lie between the collinear ones, that
demonstrates clearly the particular resummation of high-order collinear
contributions at small $x$ values in the $k_T$-factorization approach.
We also see  exellent agreement between the experimental data and
collinear approach with GRV parton densities~\cite{GRV}
at NLO approximation (the corresponding coefficient functions were taken
from the papers~\cite{Neerven}).
The NLO corrections are large and negative and decrease the $F_L$ value
by an approximate factor of 2 at $Q^2 < 10$ GeV$^2$.
\vspace*{-.3cm} 
\begin{figure}[!t]
\vspace*{7.0cm}
\begin{center}
\includegraphics{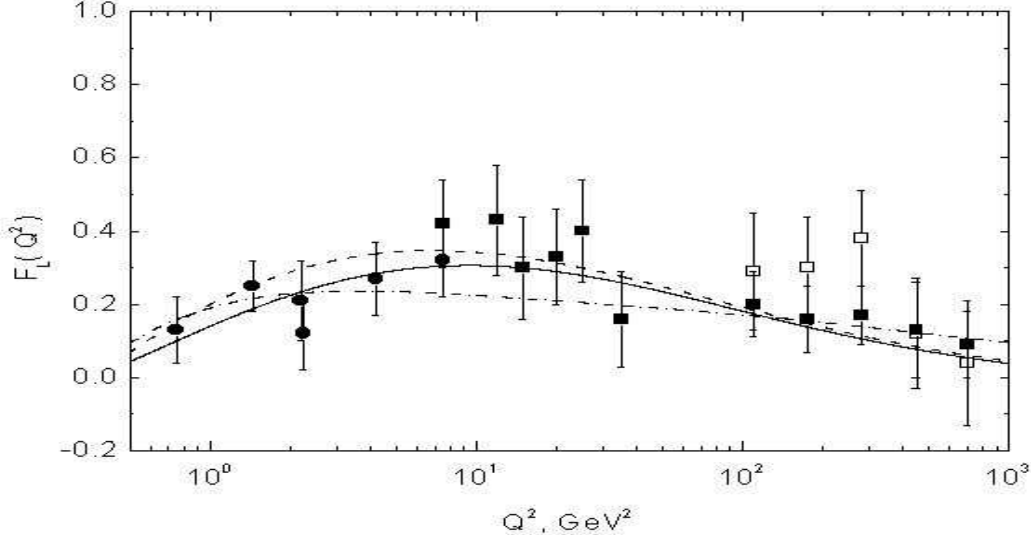}
\vspace*{.8cm}
\caption[*]{$Q^2$ dependence of $F_L(x, Q^2)$ (at fixed $W$ = 276 GeV).
The experimental points are as in Fig. 1.
Solid curve is the result of the $k_T-$factorization approach
at $\mu^2 = 127Q^2$, dashed curve - the collinear
LO calculations at $\mu^2 = 127Q^2$, dash-dotted curve - from the
$R_{world}$-parametrization.}
\vspace*{-.5cm}
\end{center}
\end{figure}
Our $k_T$-factorization
 results are in good agreement with the data for large and small
parts of the $Q^2$ range. We have, however, some disagreement
 between the data
and theoretical predictions at $Q^2 \sim 3$ GeV$^2$. The disagreement
exists in both cases: for collinear QCD approach at the LO
approximation and for $k_T$-factorization. It is possible to assume,
that the disagreement  comes from two reasons:  additional higher-twist 
contributions, which are important at low $Q^2$ values\footnote{Some part 
of 
higher-twist contributions was took into account by the  
"freezing" procedure.}, or/and  NLO QCD corrections.

   It was shown that the saturation (non-linear QCD) approaches contain
onformation of all orders in $1/Q^2$, they resum higher-twist 
contributions~\cite{Bar}. The analysis of the behaviour of the
longitudinal structure function $F_L(x, Q^2)$ in the saturatation models 
was done in Ref.~\cite{Mach}\footnote{N.Z. thanks M.V.T. Machado for
useful discussion of this problem.}. In Fig. 2 we demonstrate our 
$k_T-$factorization description of $F_L(Q^2)$ at fixed $W$ with the
unintegrated gluon distribution proposed in Ref.~\cite{LL} which
takes into account non-linear (saturation) effects.                    

Concering the NLO corrections in the $k_T$-factorization
approach a rough estimation of that can be done in the following way.
Consider firstly the BFKL approach. A popular resummation of the
NLO corrections is done in \cite{BFKLP}, which demonstrates that
the basic effect of the NLO  corrections
is the strong rise of the $\alpha_s$ argument from $Q^2$ to $Q^2_{eff} =
K \cdot Q^2$, where $K=127$, i.e. $K>>1$.

The use of the effective argument $Q^2_{eff}$ in the DGLAP approach at LO
approximation leads to results which are very close to the ones
obtained in the case of
NLO approximation: see the dot-dashed and dotted curves in Fig. 1.
Thus, we hope that the effective
argument represents the basic effect of the NLO  corrections also in the
framework of the $k_T$-factorization, which in some
 sense lies between the DGLAP and BFKL approaches.

The results obtained in the $k_T$-factorization and collinear approaches
based on $Q^2_{eff}$ argument are presented in Fig. 3.
There is very good agreement
between the experimental data and both theoretical approaches.
 
Moreover, we also present in Fig. 3 the $F_L$ results based on the
$R_{world}$-parameterization for the $R=\sigma_L/\sigma_T$ ratio (see
\cite{SLAC}) (because $F_L=F_2 R/(1+R)$), and the $F_2$ parameterization
from our previous paper \cite{KLZ1}.
The results are in good agreement with other
theoretical predictions  as well as with experimental data.

\section{Conclusion}

In the framework of $k_T$-factorization and  perturbative QCD
at LO and NLO approximations
we have analyzed recent H1 preliminary data \cite{H1rev}.

We have found very good agreement between the experimental data and
collinear NLO results.
The LO collinear and $k_T$-factorization results show a disagreement
with the H1 data in the range $2 < Q^2 < 10$ GeV$^2$ values.  
It was assumed that the disagreement
comes from the absence of additional higher-twist or/and NLO corrections.
We shown that the account of higher-twist contributions in the form
of saturation effects results in better description of the experimental 
data. The NLO corrections were modeled
by choosing large effective scale in the QCD coupling constant. 
The effective corrections significantly improve also the agreement with 
the H1 data under consideration.
\vspace{-.4cm}
\section*{Acknowledgements}
\vspace{-.2cm}
One of the authors (A.V.K.) is supported in part by
Alexander von Humboldt fellowship.
A.V.L. is supported in part by INTAS YSF-2002 grant $N^o$ 399 and
"Dinastiya" Fundation.
N.P.Z. thanks the Organizing Committee for finacial support and 
L. J\"onsson for discussion of the H1 data
\cite{H1rev}.

\end{document}